\documentclass{WileyMSP-template}

\usepackage{graphicx} 
\usepackage{xcolor}%

\begin{document}

\pagestyle{fancy}

\rhead{\includegraphics[width=2.5cm]{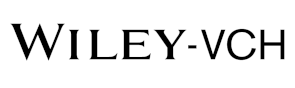}}

\title{Parallel Spatial Photonic Programming of Optoelectronic IGZO RRAM with a compact $\mu$LED Array}

\maketitle

\author{Andrew Adair}
\author{Joshua Robertson*}
\author{Andreas Tsiamis}
\author{Mohamed Awadein}
\author{Spyros Stathopoulos}
\author{Johannes Herrnsdorf}
\author{Martin D. Dawson}
\author{Themis Prodromakis}
\author{Antonio Hurtado}


\begin{affiliations}
A. Adair, Dr. J. Robertson, Dr. J. Herrnsdorf, Prof. M. D. Dawson and Prof. A. Hurtado\\
Institute of Photonics, SUPA Dept. of Physics, University of Strathclyde, Glasgow, 99 George Street, G1 1RD, UK\\
Email Address: andrew.adair@strath.ac.uk, joshua.robertson@strath.ac.uk, johannes.herrnsdorf@strath.ac.uk, m.dawson@strath.ac.uk, antonio.hurtado@strath.ac.uk

Dr. M Awadein, Dr. A Tsiamis, Dr. S. Stathopoulos, and Prof. T Prodromakis\\
Institute for Integrated Micro and Nano Systems, University of Edinburgh, Edinburgh, Alexander Crum Brown Road, EH9 3FF, UK\\
Email Address: awadein@ed.ac.uk, s.stathopoulos@ed.ac.uk, a.tsiamis@ed.ac.uk, t.prodromakis@ed.ac.uk
\end{affiliations}



\begin{abstract}

Optoelectronic resistive random-access memory elements (ORRAM) are critical emerging devices that leverage photonic technologies to bring the advantages of optical programming to traditionally electronic memristive platforms for neuromorphic computing and artificial intelligence.
In this work, a free-space optic micro-LED ($\mu$LED) array is combined with a 2-terminal oxide semiconductor ORRAM (based on IGZO\textsubscript{Rich}/IGZO active layers), to realise parallel spatial programming of form-free memristive arrays.
We report the optical and electrical programming of resistive states with potentiation/depression analysis of various stimuli parameters (pulse frequency, pulse width, pulse amplitude).
Further, we demonstrate the simultaneous photonic-electronic programming of the ORRAM with optical SET (blue 450\,nm) and electrical RESET functionality.
Persistent photocurrent is also observed and exploited as a pathway to fading memory or synaptic plasticity for temporal bit encoding.
Finally, parallel optical $\mu$LED to ORRAM channels are demonstrated to achieve the simultaneous photonic programming of multiple devices and the writing of spatial patterns across a chip of memristive IGZO devices.
This work highlights the light-enabled scalability of the optoelectronic platform and the feasibility of ORRAM to interface with spatially-multiplexed optical sources to bring neuromorphic technologies directly into applications that process and sense in the optical domain.  
\end{abstract}

\section{Introduction}\label{Intro}

Neuromorphic computing, the emulation of biological neural network structure and functionality, has emerged as a powerful tool in the wake of the rapidly expanding field of artificial intelligence (AI).
AI, and its data-intensive requirements, are exposing fundamental bottlenecks in conventional Von-Neumann computing architectures, creating inefficiencies that scale to excessive energy consumption and latency during data processing \cite{Xia2019}.
The neuromorphic computing paradigm however, aims to design systems from the ground-up to natively achieve brain-inspired neural network functionality that enhances the parallelism and computational efficiency of hardware processing architectures. 

Memory resistors (memristors) or resistive random-access memory (RRAM), are one of the most widely investigated key-enabling electronic candidates for neuromorphic computing hardware \cite{Xiao2023}.
These electronic devices allow for the programming of their internal resistive state with application of voltage and/or current stimuli, and exhibit a direct dependence on historic inputs with the capability to even maintain states after power loss \cite{Chua1971}.
Inherently capable of remembering, RRAM devices have demonstrated both long-term and short-term memory, realised through non-volatile and volatile resistive switching mechanisms, respectively \cite{Zhao2025}.
The switching mechanisms and material systems are the primary means of categorising RRAM devices, with approaches including ion migration \cite{Yang2012,Waser2009,Serb2016,Papadopoulos2022}, electronic trapping \cite{Lu2019}, and phase transitions \cite{Boybat2018}.
The programmable nature of memristive devices, alongside their low power consumption, compatibility with Complementary Metal-Oxide-Semiconductor (CMOS) foundries \cite{Tsiamis2026}, high switching speeds and scalability, fuels the motivation behind their neuromorphic deployment.
RRAM devices having since demonstrated synaptic \cite{Ji2025} functionality and even full neuronal operation \cite{Ji2025,Park2022,Zhao2025}. 

The integration of electronic RRAM with photonic technologies aims to marry the advantages of both platforms, granting access to high bandwidth, low cross talk and parallelism of optical inputs for optical programming with reduced joule heating \cite{Zhou2021}.
The optoelectronic RRAM platform, known as ORRAM, also positions itself well to exploit additional degrees of freedom (3D-space, wavelength, polarisation, etc.) that can be deployed to multiplex operations for enhanced dynamic control of the platform (for advanced learning techniques), and direct implementation in optical sensory systems \cite{Zhou2021,Jaafar2019}.
Further, the deployment of ORRAM for preprocessing in artificial vision systems has been proposed, enabling optical to electrical conversion with additional functionality (noise reduction or contrast enhancement) \cite{Mao2019}, and in-sensor reservoir computing, leveraging on the short-term plasticity of the optical-encoded inputs \cite{Sun2023}. 

Material systems including metal oxides \cite{Ungureanu2012,Zhou2019,Wang2023}, semiconductor oxides \cite{Yang2024,Mehonic2017,Jaafar2023}, 2D materials \cite{Jaafar2019}, 1D structures \cite{Shan2022} and organics \cite{Jaafar2024}, have all been used to demonstrate ORRAM devices with various levels of optical and electrical programmability (see \cite{Pereira2023} for a review).
Among these, indium gallium zinc oxide (IGZO)-based ORRAM have gained particular attention due to their persistent photoconductivity and high light absorption.
Recent studies have demonstrated IGZO ORRAM devices with both optical SET and optical RESET capability using blue and red light irradiation \cite{Hu2021}, enhanced sensitivities across a range of visible and UV wavelengths \cite{Pereira2024,Wu2024}, and even combined optical operation with an oscillator neuron for the artificial reconstruction of retina neural circuitry \cite{Wu2020}.
These reports highlight the appropriateness of the IGZO ORRAM platform for neuromorphic systems with demonstrations of short- and long-term memory, spike-timing dependent plasticity and pulse pair facilitation, with all optical or hybrid optoelectronic stimulation.

In this work, we leverage the parallelism of a photonic micro-LED ($\mu$LED) array source to develop a novel ORRAM demonstrator platform that can controllably program multiple individual IGZO devices on-chip simultaneously.
Specifically, dual-stack devices of IGZO\textsubscript{Rich}/IGZO active layers are illuminated by a compact, efficient $\mu$LED platform featuring gallium nitride (GaN)-based light-emitting diodes (operating at the wavelength of 450\,nm \cite{Herrnsdorf2015}), to demonstrate the efficient scaling of neuromorphic ORRAM operations with spatially-multiplexed optical platforms.
The $\mu$LED array system features light-emitting diodes with active areas less than 100\,$\mu m^2$, bump-bonded directly to CMOS for crucial individual pixel control \cite{Rae2008} and high modulation rate capability \cite{Ferreira2016}, towards high brightness display \cite{Herrnsdorf2015} and optical wireless communication \cite{McKendry2012, Ferreira2016} applications.
Here it is deployed as an advanced programmable source for an array of in-house developed IGZO ORRAM.
Firstly, we investigate the potentiation and depression of electronic states of IGZO ORRAM under electrical, optical and optoelectronic stimuli, before leveraging the persistent photocurrent to realise light-driven temporal bit pattern encoding through a short-term fading memory (synaptic plasticity), and finally a multi-device spatial photonic pattern encoding through parallel one-to-one $\mu$LED-ORRAM channels.

\section{Results}

\begin{figure}
\centering
\includegraphics[width=0.85\linewidth]{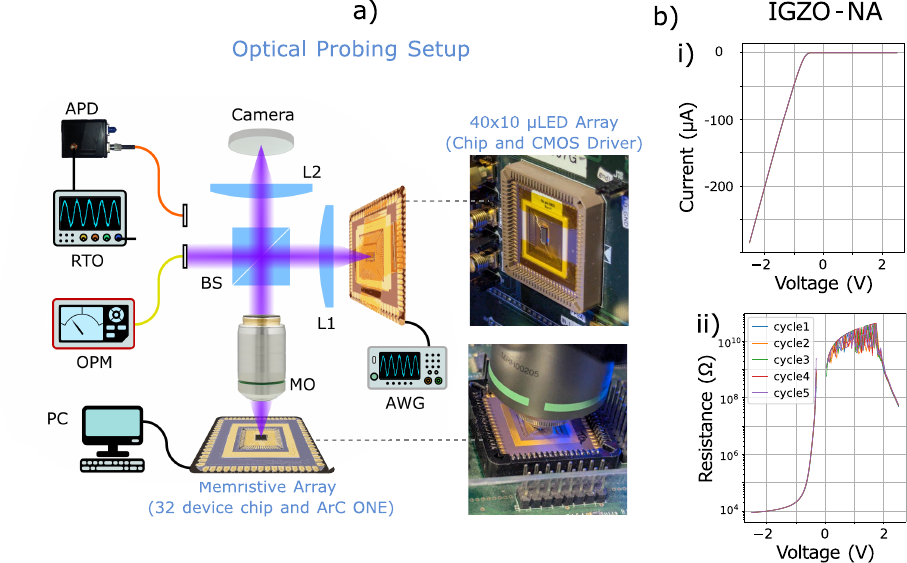}
\caption{\label{fig:Fig1}
    (a) Experimental setup for optical probing of IGZO ORRAM devices. Blue light from a $\mu$LED array is coupled into ORRAM devices using free-space optics. APD - Avalanche photodetector, RTO - Real-time oscilloscope, OPM - Optical power meter, L1/L2 - Plano-convex lens, BS - 70R:30T Beamsplitter, MO - Microscope objective, AWG - Arbitrary waveform generator. 
    (b) Characteristic (i) I-V and (ii) R-V curves (Current and Log Resistance) for IGZO-NA. Curves cycled 5 times between -2.5\,V and +2.5\,V.}
\end{figure}

The response of the fabricated ORRAM devices to electrical and optical excitation was characterised using the optical probe station shown in Fig.\,\ref{fig:Fig1}(a).
Blue, 450\,nm wavelength, light from a $\mu$LED array was focused onto wire-bonded chips containing 32 stand-alone IGZO ORRAM devices (see setup discussion in Section\,\ref{sec:Setup}).
The ORRAM chips, one of a non-annealed IGZO stack (IGZO-NA), and another with an additional temperature annealing (IGZO-A)(see Section\,\ref{sec:fabrication}), were measured using an ArC ONE instrumentation platform.
The electroforming-free current-voltage (I-V) and resistance-voltage (R-V) characteristics of the IGZO-NA ORRAM devices, are plotted in Figs.\,\ref{fig:Fig1}(b)(i) \& (ii).
The applied pulse voltage was swept between -2.5\,V and +2.5\,V across 5 cycles.
The electrical characteristics of the IGZO-A sample are provided in supporting information (see Fig.\,S1).
Figure\,\ref{fig:Fig1}(b)(i) shows a rectifying behaviour present within the IGZO-NA devices where the current depends on the polarity of voltage applied.
Figure\,\ref{fig:Fig1}(b)(ii) reveals further that the IGZO-NA ORRAM devices have an ohmic linear relationship for negative polarity voltages, and a Schottky contact-induced region of reduced conduction for positive voltage.
Here, the ORRAM device reached resistances of up to 10$^{10}\,\Omega$, exceeding the maximum ammeter resolution of ArC ONE platform.
This resulted in noisy resistance measurements from 0 to 2\,V.
This setup limitation can be directly addressed through the annealing of the IGZO sample (see supporting information, Fig.\,S1), reducing the intrinsic resistance by several orders of magnitude. 

\subsection{Electrical Characterisation}\label{Elec_character}

\begin{figure}
\centering
\includegraphics[width=1\linewidth]{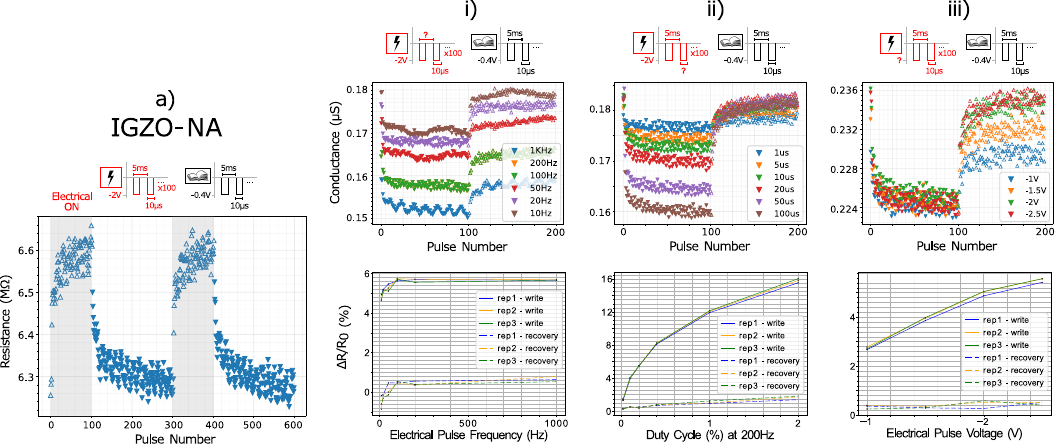}
\caption{\label{fig:Fig2}
    (a) Electrical characterisation of the potentiation-depression property of IGZO-NA with: 
    i) varying pulse frequency, 
    ii) varying pulse width, 
    iii) varying pulse amplitude.
    (Top) conductance ($\mu S$) and (Bottom) relative resistance change (\%) of the devices. 
    Measurements are taken over 3 cycles, with 2 cycles shown in (a), and only the first cycle shown in conductance plots (top). 
    The constant parameters are read voltage = -0.4\,V, read interval = 50\,ms, WRITE voltage = -2\,V, pulse width = 10\,us, pulse interval = 5\,ms, pulse count = 100.
    }
\end{figure}

A series of characterisations were performed on both IGZO-NA and IGZO-A samples to assess the form-free performance when subject to electrical excitation. 
The potentiation and depression properties of the two IGZO devices were tested by varying three parameters of the applied electrical stimuli: pulse frequency, pulse width, and pulse voltage.
For this characterisation, 100 excitation pulses were applied, followed by 200 read pulses to observe the recovery of the device (without external stimulation).
Negative polarity pulses were selected for excitation as the ohmic side of the R-V plots (shown in Fig.\,\ref{fig:Fig1}(b)(ii) \& Fig. S1(b) in supporting information) achieved a larger change in resistance.
The electrical characterisation of the IGZO-NA devices are shown in Fig.\,\ref{fig:Fig2}, with IGZO-A shown in supporting information (see Fig.\,S2).
For IGZO-NA, the resistance was read using 10\,$\mu$s long pulses of -0.4\,V at an interval of 5\,ms.
A consistent set of starting excitation parameters (-2\,V, 10\,$\mu$s width, 5\,ms interval/200\,Hz frequency) were used across the input pulse studies.
In total three consecutive cycles of potentiation and depression were measured and analysed.   

Figure\,\ref{fig:Fig2}(a)(i) demonstrates the potentiation and depression of the IGZO-NA sample when the WRITE frequency was increased from 10\,Hz to 1\,kHz.
Similarly, in Fig.\,\ref{fig:Fig2}(a)(ii) the WRITE width was increased from 1\,$\mu$s to 100\,$\mu$s, and in Fig.\,\ref{fig:Fig2}(a)(iii) the WRITE voltage was increased from -1\,V to -2.5\,V.
The measurements show that all electrical input pulses produced an increase in device resistance.
Following the removal of the excitation, the devices demonstrate a volatile behaviour, initially recovering rapidly (with a decreasing resistance following approximately 20 consecutive read pulses), before slowly plateauing for the remainder of the recovery measurement.
The conductance of the device, alongside the relative resistance change (\%) during a potentiation-depression cycle, is shown for each analysis case.
The relative resistance change is measured as the difference in resistance from the start to the end of the potentiation ($\Delta$R), divided by the initial resistance (R$_{0}$), utilising a moving average to improve readout noise.
The plot demonstrates both the relative resistance change of the WRITE pulses (solid lines), and the relative resistance change of the recovery (dashed line) across 3 cycles. 

The characterisation results reveal that when considering the electrical WRITE operation, little effect is observed when changing input frequency/interval, with a stable relative resistance change across 2 orders of magnitude.
However, in both the study of WRITE pulse width and voltage, a measurable influence on the relative resistance change is observed, with large width and large amplitude inputs driving the resistive state of the device up most significantly.
Pulse width and voltage can therefore be leveraged to controllable program the resistive state of the device during an electrical WRITE process.
Without stimulation, the ORRAM demonstrates a fast recovery $\sim$\,0.1\,s, almost completely resetting to R$_{0}$ irrespective of the electrical WRITE properties.
This indicates these ORRAM devices are highly volatile with only short-term memory when operating with electrical stimuli.
These electrical WRITE characteristics were found to be consistent across IGZO samples (see supporting information for IGZO-A).

\subsection{Optical Characterisation}\label{Optical_character}

\begin{figure}
\centering
\includegraphics[width=1\linewidth]{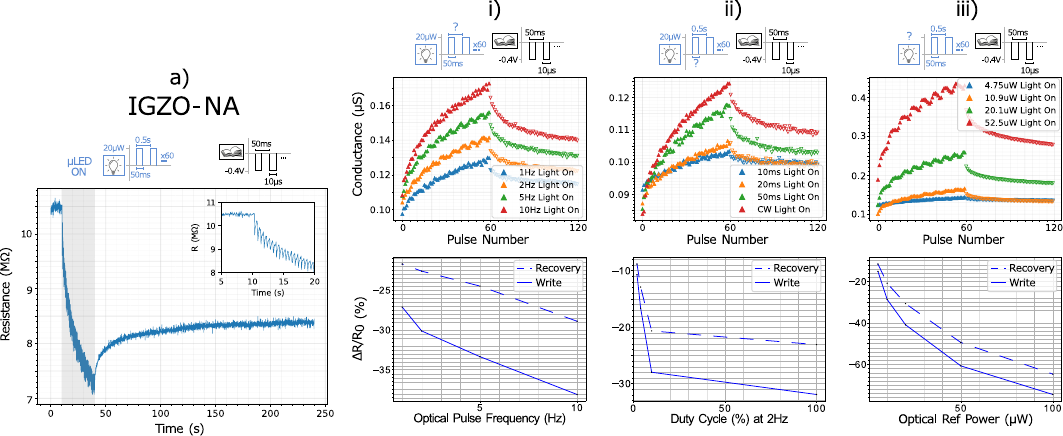}
\caption{\label{fig:Fig3}
    Optical characterisation of the potentiation-depression property of IGZO-NA with: 
    i) varying pulse frequency, 
    ii) varying pulse width, 
    iii) varying optical power.
    (Top) conductance ($\mu S$) and (Bottom) relative resistance change (\%) of the devices. 
    Measurements are taken over a single cycle. 
    The constant parameters are read voltage = -0.4\,V, read interval = 50\,ms, optical power = 106\,$\mu W$, pulse width = 50\,ms, pulse interval = 500\,ms, pulse count = 60.}
\end{figure}

Following electrical testing, the key optical characteristics of the IGZO ORRAM devices were measured using a series of pulsed stimuli. 
Similar to the previous characterisation, but now deploying the $\mu$LED array system described in Fig. \ref{fig:Fig1}, the light-induced programming of ORRAM resistance was performed with a single $\mu$LED source. 
For this characterisation, 60 optical excitation pulses were applied, followed by 400 read pulses to observe the recovery of the device. 
The optical excitation parameters were again varied during potentiation and depression measurements, with optical pulse frequencies from 1 to 10\,Hz, optical pulse widths from 10\,ms to continuous wave (CW), and optical reference powers from 4.75 up to 106\,$\mu W$. 
Here, with a beam spot diameter of 45\,$\mu m$, reference powers of 4.75, 10.9, 20.1, 52.5 and 106\,$\mu W$ produced calculated optical power densities of 53.1, 113, 220, 560, and 1156\,$mWcm^{-2}$. 
The measurement of IGZO-NA (shown in Fig.\,\ref{fig:Fig3}\,(a)) used electrical read pulses of 10\,$\mu$s width and -0.4\,V at an interval of 50\,ms. 
The optical characteristics of IGZO-A are presented in supporting information (see Fig. S3). 
Again, a consistent set of starting excitation parameters were used (2\,Hz optical pulse frequency, 50\,ms pulse width, and an optical reference power of 106\,$\mu W$).

Figure\,\ref{fig:Fig3}\,(a) reveals the potentiation and depression of the IGZO-NA sample where, under illumination, the resistance of the ORRAM device decreases from its initial state (R$_{0}$). 
This corresponds to an optical WRITE operation that is inverse to that of the electrical WRITE presented previously in Section \ref{Elec_character}. 
The analysis of the conductance and relative resistance change is shown for each varying parameter (Figs.\,\ref{fig:Fig3}\,(a)(i-iii)). 
Figures\,\ref{fig:Fig3}\,(a)(i-iii) show that an increase in optical frequency, optical pulse width and optical power, each result in a relative resistance change, superior to that observed with electrical stimuli. 
The variation of the optical power parameter has the largest influence on the WRITE operation for IGZO-NA (shown in Fig.\,\ref{fig:Fig3}\,(a)(iii)), where varying the reference power from 4.75\,$\mu W$ to 106\,$\mu W$ alters the relative resistance change by over 55\,\%, or a conductance change of over 0.2\,$\mu S$. 
Additionally, unlike electrical stimulation, the frequency of the optical input directly influences the final state of the ORRAM devices. 
Therefore, in order to controllably write levels in the tested ORRAM devices, all 3 WRITE parameters should be considered. 
When optical stimuli are removed, the device slowly recovered to a resistive state below R$_{0}$, failing to fully recover within the 250\,s of the measurement. 
This can be observed in the bottom plots of Figs.\,\ref{fig:Fig3}\,(a)(i-iii), where the 250\,s recovery typically increases the relative resistance by 10 to 15\%. 
Recovery following optical stimulation is therefore much slower than the $\sim$\,0.1\,s of electrical stimulation, enabling optical WRITE operations to encode volatile fading memories that exist for longer.
Overall, the optical characteristics of IGZO-NA reveal a good optical programming capability that can achieve a change in device conductance of up to 0.2\,$\mu S$ with optical illumination from a single $\mu$LED source.
Optical excitation of the annealed IGZO samples (IGZO-A) was also investigated, revealing lower overall optical sensitivity, as discussed in supporting information (see Fig.\,S3(a)).

\subsection{Optical-SET/Electrical-RESET Characterisation}

\begin{figure}[t!]
\centering
\includegraphics[width=1\linewidth]{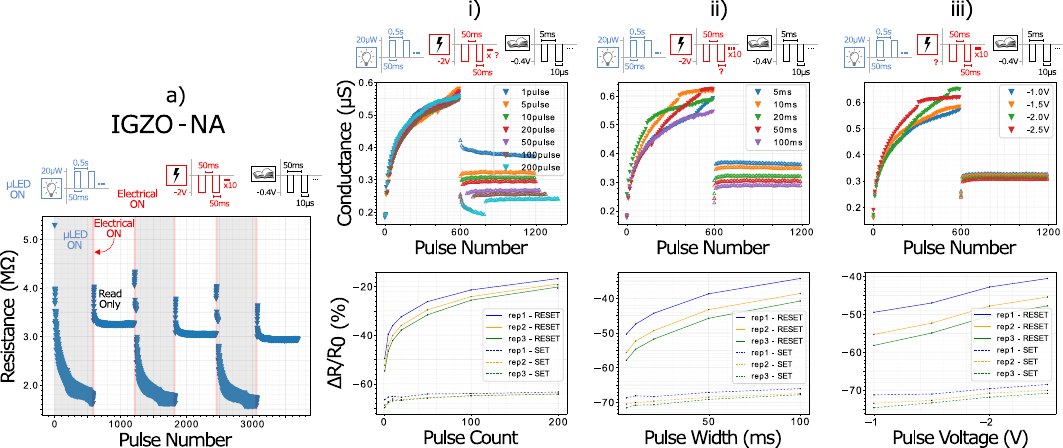}
\caption{\label{fig:Fig4}
    Optical-SET electrical-RESET characterisation of IGZO-NA, with: 
    i) varying pulse count, 
    ii) varying pulse width, 
    iii) varying pulse amplitude.
    (Top) conductance ($\mu$S) and (Bottom) relative resistance change (\%) of the devices. 
    Measurements are taken over 3 cycles, with all 3 cycles shown in (a), and only the first cycle shown in conductance plots (top). 
    The constant parameters are read voltage = -0.4\,V, read interval = 50\,ms, optical power = 106\,$\mu W$, pulse width = 50\,ms, pulse interval = 500\,ms, pulse count = 60.}
\end{figure}

Following the electrical and optical characterisations, the optoelectronic operation of the ORRAM was tested. 
Specifically, optical and electrical WRITE pulses were combined to program the final resistive state of a IGZO-NA device. 
Here the optical and electrical WRITE pulses counteract one another, allowing to SET a resistive state with pulses of light, and RESET the resistive state with electrical inputs. 
For this characterisation, 3 parameters of the electrical RESET pulses were varied following optical excitation: electrical pulse count, pulse width, and pulse voltage. 
The optical-SET electrical-RESET was repeated for 3 cycles, as shown in Fig.\,\ref{fig:Fig4}(a). 
The IGZO-NA device was SET with 60 optical pulses from a single $\mu$LED, using a fixed 2\,Hz frequency, 50\,ms pulses width, 106\,$\mu$W reference optical power. 
A consistent set of starting RESET parameters were used (10 electrical pulses, at 50\,ms intervals, with -2\,V, and 50\,ms width). 
The resistive state of the device was read 600 times (-0.4\,V, 10\,$\mu s$ width, 20\,Hz frequency) during the optical SET, and 600 times after the electrical RESET sequence. 

The analysis of the optical-SET electrical-RESET operation is provided in Figs.\,\ref{fig:Fig4}\,(a)(i)-(iii). 
The measured conductance of the device for the first cycle is shown in the top plots, and the relative resistance change is shown in the bottom plots. 
The relative resistance change is measured for both the end of the optical SET operation (dotted-line) and the end of the electrical RESET operation (solid-line), with the initial resistance (R$_{0}$) taken as the state prior to optical illumination. 
Figures\,\ref{fig:Fig4}\,(a)(i)-(iii) reveal that increasing each RESET pulse parameter (number of RESET pulses, pulse width, pulse voltage) results in an improved recovery, pushing the resistance closer to R$_{0}$. 
The RESET parameter with the largest influence on resistance was pulse count, with a $\sim$\,35\,\% improvement obtained when increasing the number of electrical input pulses from 1 to 200 (as seen in Fig.\,\ref{fig:Fig4}\,(i)). 
Using 200 electrical pulses, the IGZO-NA device was controllably reset to 80\% of the initial resistance value in only 10\,s, significantly faster than the recovery observed in Section \ref{Optical_character}. 
The electrical RESET did not fully recover the initial resistance for the parameters used in this characterisation, however, this could be achieved by further fine-tuning the SET and RESET parameters. 
The optical-SET electrical-RESET system enables these optoelectronic IGZO ORRAM devices to be used as programmable memory devices with a suite of controllable SET parameters and RESET parameters. 
Optoelectronic operation can be applied to counteract the slow recovery induced by the optical excitation of the system, clearing fading memory, or be leveraged to enable simultaneous multi-modal programming where 2 variables fuse to create resistive memory states. 

\subsection{Temporal Bit Pattern Encoding}\label{Temp Pattern Encoding}

To highlight the potential deployment of the ORRAM and its programmable fading memory (synaptic plasticity), we performed a photonic temporal 4-bit encoding demonstration with a single IGZO-NA device. 
The objective of this demonstration was to encode a time-multiplexed bit pattern using a sequence of optical SET operations, and using the readout of the final resistive state, distinguish the input bit sequence from others of the set. 
To achieve this, the resistance of the device is first RESET using 400 electrical pulses of -2\,V amplitude, 100\,ms width, with a 50\,ms interval. 
The IGZO-NA device then receives one of 16 optically-encoded bit patterns, where each bit ('0' or '1') corresponds to optical inputs from a $\mu$LED of the array. 
Bits valued '1' illuminate the ORRAM device with 10 optical pulses at 2\,Hz, 50\,ms pulse width and 106\,$\mu W$ optical reference power. 
Bits values '0' result in no optical signal. A single electrical RESET pulse (-2\,V, 5\,ms width and 50\,ms interval) is applied between the sequential bits to raise the resistance of the device and prevent saturation from multiple optical inputs. 
The resistive state of the device is measured using constant electrical readout pulses (-0.4\,V, 50\,ms width, 50\,ms interval).

\begin{figure}
\centering
\includegraphics[width=0.8\linewidth]{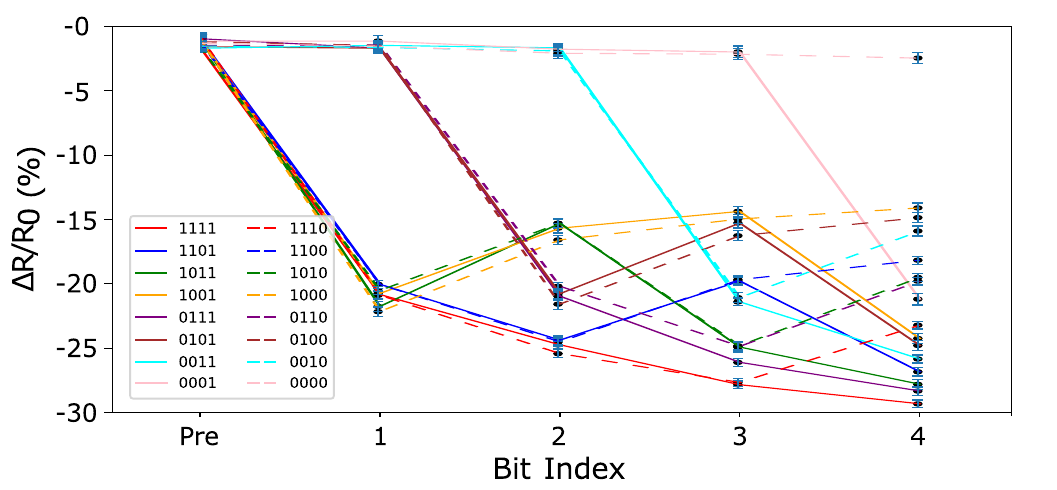}
\caption{\label{fig:Fig5}
    Optical temporal bit pattern encoding with a single IGZO-NA ORRAM device exploiting fading memory and synaptic plasticity. Following the application of all 4 bits, the final resistive state readout can be used to distinguish the full input sequence. The constant parameters are read voltage = -0.4\,V, read interval = 50\,ms, pulse width = 50\,ms, pulse interval = 500\,ms, optical power = 106\,$\mu W$, pulse count = 60.}
\end{figure}

The results of the temporal pattern encoding are shown in Fig.\,\ref{fig:Fig5}. 
Figure\,\ref{fig:Fig5} plots the final resistive state of the IGZO-NA device following the encoding of each bit of the pattern. 
After the application of all four bits, the full temporal pattern is represented by a final readout value. 
This transformation is achieved by integrating the resistive state of each bit with the fading memory of the previous bit. 
This effect is clearly observed in Fig.\,\ref{fig:Fig5} after the encoding of the second bit. 
Here, four distinct clusters of states emerge that each directly depend on the high or low resistance memory of the encoded first bit. 
The high reproducibility of the encoding can be seen through the early overlapping of traces with the same initial bits. 
Error bars show the standard deviation of the resistive state over 150 electrical readout pulses prior to optical injection ($\pm$\,0.35\%). 
Following the encoding of all four bits, the system produces up to 15 distinct resistive states allowing for the good distinguishing of photonic temporal patterns. 
Overall this demonstration of temporal bit pattern encoding realised separable resistive states that classify 4-bit patterns with an accuracy of 87.5\% (14 of 16 patterns). 
Given further optimisation and tuning of the SET and RESET procedure, this system could realise error-free 4-bit temporal encoding, with temporal patterns '0110' and '1010' currently overlapping within the error of the electrical readout. 
Several reasons can contribute to the converging of the two different temporal pattern states. 
Undesired fluctuations in the ArC One ammeter lead to an increased standard deviation of read noise in the system, increasing our margin for resistive state overlap. 
Moreover, the balance of optical SET and electrical RESET could be further optimised. 
Currently, there is a non-uniform spread of resistive states within the range of the ORRAM device. 
In total, 15 patterns produced final readouts in the range of -15\% to -30\%, with only pattern '0000' producing a resistive state between 0 and -15\%. 
This signifies that the parameters of the optical SET were over-tuned (saturating the states towards lower resistance values) or that the parameters of the electrical RESET were under-tuned (failing to push the resistance of the ORRAM device higher). 
Given the mean read error of $\pm$\,0.35\%, and the currently observed resistive range (0 to -30\%), the ORRAM device could host up to 85 distinguishable states, potentially allowing for neuromorphic temporal encoding of up to 6 bits with these ORRAM devices.

\subsection{Parallel ORRAM Programming with a $\mu$LED Array}

To increase the scale of the ORRAM optical programming, we finally demonstrate the parallel operation of the photonic system. 
To that end, we demonstrate 4 optical SET operations on 4 IGZO-NA devices in parallel. 
Specifically, four adjacent non-annealed IGZO devices (devices 20-26 in the fabricated array) are spatially coupled to their own individual-modulated $\mu$LEDs, deploying the same setup described in Fig.\,\ref{fig:Fig1}\,(a). 
For this demonstration four IGZO-NA ORRAM devices were successfully programmed, however, larger array-to-array coupling configurations will be possible given the pitch between ORRAM devices and $\mu$LEDs is used in combination with the correct optical magnification. 
In the future, the optical magnification and $\mu$LED pitch could inform ORRAM array design to enable full array-to-array coupling.  

\begin{figure}[t!]
\centering
\includegraphics[width=0.65\linewidth]{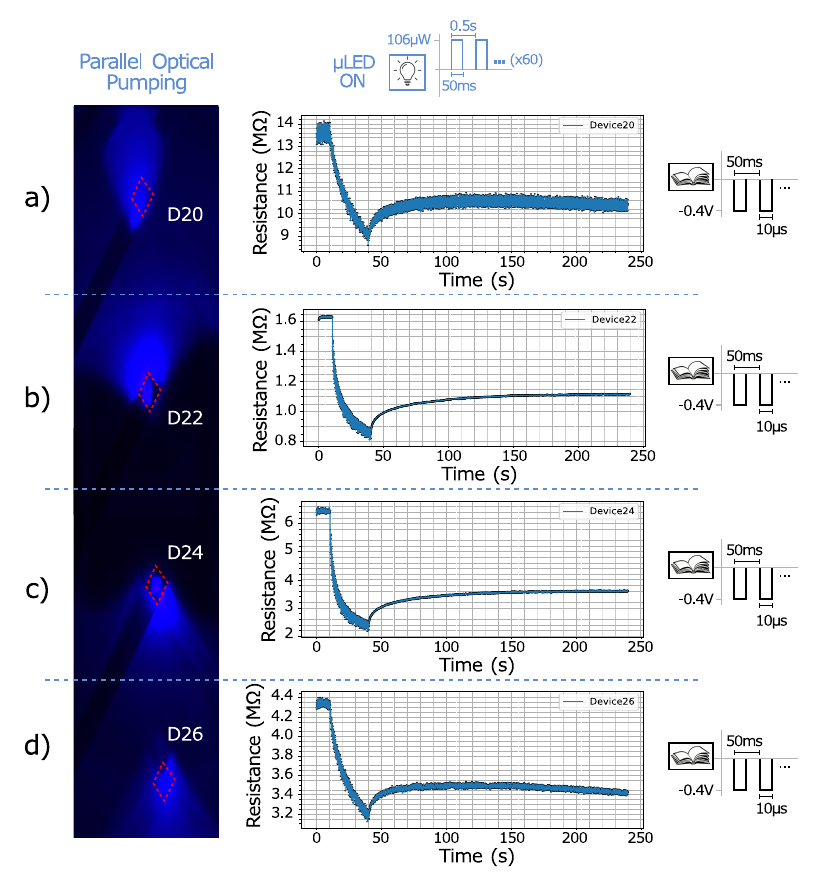}
\caption{\label{fig:Fig6}
    Demonstration of parallel optical injection of four $\mu$LEDs devices into four adjacent form-free ORRAM devices. Each optical illumination results in the potentiation of each ORRAM device. The constant parameters are read voltage = -0.4\,V, read interval = 50\,ms, pulse width = 50\,ms, pulse interval = 500\,ms, optical power = 106\,$\mu W$, pulse count = 60.
    }
\end{figure}

Figure\,\ref{fig:Fig6} shows stitched microscope camera images of the ORRAM array under illumination from multiple $\mu$LEDs, and the resistance drop of each IGZO-NA device over time. 
The optical stimulation consisted of 60 pulses at 2\,Hz, with pulse widths of 50\,ms and a reference power of 106\,$\mu$W. 
This optical input was sent in parallel by all four $\mu$LEDs. 
Short 10\,$\mu$s read pulses of -0.4\,V were used to monitor the resistive state of the four devices at 20\,Hz. 
Figures\,\ref{fig:Fig6}\,(a)-(d) successfully demonstrate the parallel optically-driven potentiation of each device when subject to input light injection. 
Significant resistance changes were achieved in each device, reaching -33.3\%, -46.9\%, -61.5\% and -25.6\%, for devices 20-26, respectively. 
The disparity in the resistance change between devices can be explained by the limitations of the setup. Currently the outer two $\mu$LED devices (20 and 26) are incident on the lenses off-axis, resulting in distortion and aberration (as seen in the microscope image). 
Further, this leads directly to optical power loss, with the same devices coupling 75\% less power out the MO onto the ORRAM array, than devices 22 and 24. 
The reduced optical power and the distortion of the optical spot result in less coupling into the ORRAM devices and resistance reduced programmability. 


\begin{figure}
\centering
\includegraphics[width=1\linewidth]{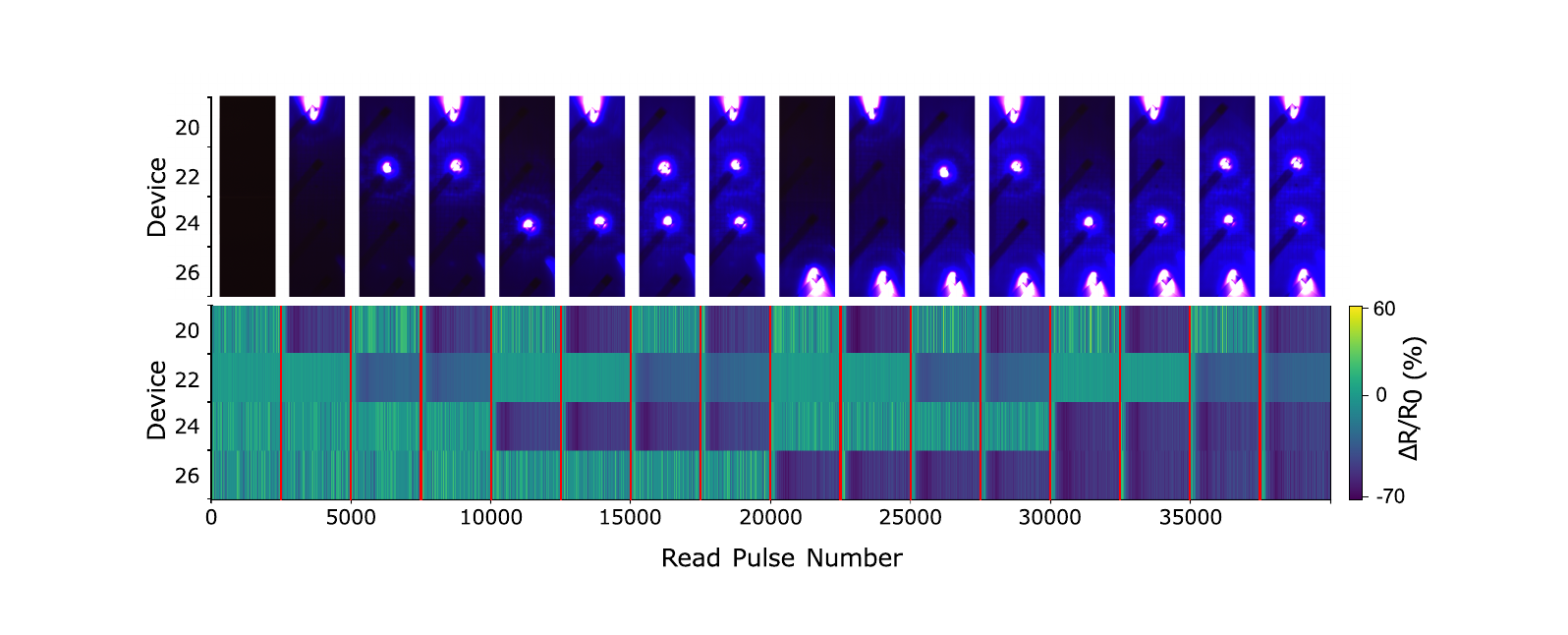}
\caption{\label{fig:Fig7}
    Parallel programming of four ORRAM devices with 16 four-bit spatial patterns. Microscope camera images and colourmaps show the illumination and potentiation of each device. Optical programming is performed via four parallel $\mu$LEDs to ORRAM channels, with 400 (-2\,V, 100\,ms width, 50\,ms interval) electrical RESET pulses (red lines) between pattern programming cycles. The constant parameters are read voltage = -0.4\,V, read interval = 50\,ms, pulse width = 50\,ms, pulse interval = 500\,ms, optical power = 106\,$\mu W$, pulse count = 60.
    }
\end{figure}

To highlight that each ORRAM-$\mu$LED pair was isolated and that cross-talk between IGZO-NA devices could be avoided during optical programming, we tested the system with different spatial light patterns from the $\mu$LED array. 
In total 16 spatial $\mu$LED patterns were projected onto the ORRAM array. 
Before each pattern was optically programmed, the resistance of the ORRAM devices was RESET with electrical pulses (x400, -2 V amplitude, 100 ms width, 50 ms interval). 
During this task sequentially projected spatial patterns were intentionally uncoupled, to avoid any influence of fading memory (represented by the red lines in Fig.\,\ref{fig:Fig7}). 
Similar to Fig.\,\ref{fig:Fig6}, spatial positions with an active $\mu$LED were programmed to inject 60 optical pulses at 2\,Hz, with pulse widths of 50\,ms and a reference power of 106\,$\mu$W. 

The resistive state of each IGZO-NA device (20-26) is plotted in the colourmap of Fig.\,\ref{fig:Fig7}. 
Potentiation corresponds to dark-blue map segments, with the resting noisy states shown in the first panel where no $\mu$LEDs are active. 
The results demonstrate that each ORRAM device potentiates correctly to the injection of all spatial $\mu$LED patterns. 
Further, the results show that little optical cross-talk between neighbouring ORRAM devices is produced, given drops in resistance only occur for devices exposed to optical input. 
This can be attributed to the small spot size of the focused light beams and the pitch of the ORRAM array. 
The selection of optical magnification is therefore a critical parameter when considering future system designs. 

This result demonstrates that multiple optoelectronic IGZO-NA ORRAM devices can be optically programmed simultaneously with a compact, efficient array of $\mu$LED sources. 
This highlights the feasibility and effectiveness of deploying spatial-multiplexing for the upscaling of optically sensitive memristive architectures. With this approach, kernels, images, matrices, and data could all be spatially written into resistive states with short optical flashes. 
Further, similar to Section\,\ref{Temp Pattern Encoding}, accessing the intrinsic fading memory of multiple ORRAM devices in parallel would massively increase the bandwidth of timeseries operations while also enabling change-point analysis of different two dimensional modalities. 
Overall, the parallel programming of ORRAM devices with optics exacerbates the deployment of light-based technologies, allowing the optoelectronic system to better benefit from speed and energy properties of photonics that make its application so appealing.

\section{Methods}\label{Method}

\subsection{Fabrication and Materials}\label{sec:fabrication}

The proposed optoelectronic ORRAM is a two-terminal metal-insulator-metal architecture, with the devices configured as 32 stand-alone cells per chip. 
Each chip comprises ORRAM of a single (active area) dimension, with available dimensions ranging from 60 x 60 \textmu m\textsuperscript{2} down to 1 \textmu m\textsuperscript{2}. 
The devices were fabricated on 150\,mm diameter, p-type, \textless 100\textgreater\ orientation silicon wafers (Si-Mat), where a bottom insulator of 200\,nm silicon dioxide (SiO\textsubscript{2}) was first grown by thermal oxidation. 

Three photolithographic steps were utilised to pattern the bottom electrode (BE), active layer and top electrode (TE) of the ORRAM, while a final photolithography defined the pattern for a TE bond-pad adhesion enhancement layer. 
Briefly, for each step, the wafers were first treated with hexamethyldisilazane (HMDS) vapour for 10 min to enhance photoresist adhesion and spin-coated (POLOS SPIN200i) with 3\,$\mu m$ thick AZ nLOF 2035 negative photoresist. 
This was followed by photoresist soft-bake on a hotplate at 105 $^{\circ}$C for 1 min, contact ultraviolet (UV) exposure (120\,$mJcm^{-2}$) on a Karl Suss MA8 mask aligner, post-exposure bake at 105\,$^{\circ}$C for 1 min and photoresist development in AZ 726-MIF for 90\,s. 
The wafers were then rinsed with deionised water (DIW) and dried with nitrogen (N\textsubscript{2}). Finally, a 2 min photoresist descum by oxygen (O\textsubscript{2}) plasma was performed on an Electrotech 508 barrel asher (350\,W power, 800\,mT pressure).

\begin{figure}[t!]
\centering
\includegraphics[width=1\linewidth]{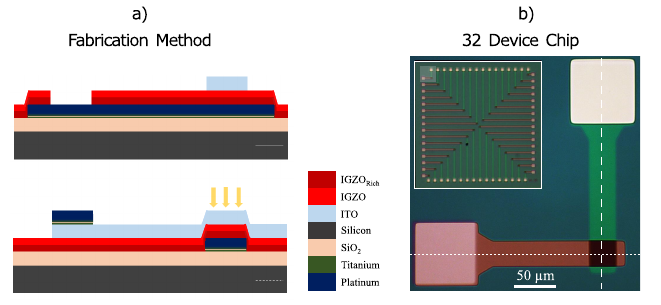}
\caption{\label{fig:Fig8}
    Design and fabrication of IGZO ORRAM devices. a) Cross-sectional diagram of the fabrication method and stack structure. b) Optical microscopy image of a single ORRAM device (bottom-right) and the full 32-device configuration on chip (top-left).}
\end{figure}

Each of the photolithographic steps was followed by a deposition process, thin-film patterning by lift-off in N-Methyl-2-pyrrolidone (NMP) at 65\,$^{\circ}$C, and finally wafer rinsing with isopropyl alcohol (IPA) and DIW, and drying with N\textsubscript{2}. 
5\,nm of titanium (Ti) adhesion layer and 12 nm of platinum (Pt) were deposited by evaporation (Angstrom Engineering) for the BE. 
The active layer is comprised of dual-stack of 12.5\,nm of oxygen-rich indium-gallium-zinc oxide (IGZO\textsubscript{Rich}) and 12.5\,nm of IGZO. 
The bottom IGZO\textsubscript{Rich} layer was deposited by radio frequency (RF) sputtering (Angstrom Engineering) from an 99.99\% InGaZnO\textsubscript{4} target in an argon (Ar) rich plasma (20 sccm gas flow), at 110\,W power and 4\,mT process pressure, with O\textsubscript{2} as the reactant gas (3\,sccm gas flow). 
The top IGZO layer was RF sputtered from the same target in an Ar plasma at 110\,W power and 4\,mT process pressure. 
IGZO-A devices were additionally annealed at 350\,$^{\circ}$C for 60 min, post active-layer deposition and patterning, using the same sputtering chamber (at 50\,mT pressure and with a constant 20\,sscm O\textsubscript{2} flow). 
20\,nm of indium tin oxide (ITO) was then deposited for the TE by RF sputtering, from a 99.99\% In\textsubscript{2}O\textsubscript{3}/SnO\textsubscript{2} (90/10 wt \%) target, at 150\,$^{\circ}$C substrate temperature, 50\,W power and 1 mT process pressure. 
5\,nm Ti and 50\,nm Pt were then evaporated (Angstrom Engineering), allowing for a top-up of the TE electrode bond-pad. 
Representative cross-sectional schematics of the devices can be seen in Fig. \ref{fig:Fig8}\,(a), while an optical microscopy image of a fabricated 30 x 30 \textmu m\textsuperscript{2} ORRAM is presented in Fig. \ref{fig:Fig8}\,(b). 
The wafers were finally diced to 3 x 3\,mm chips (see inset of Fig. \ref{fig:Fig8}\,(b)) using a DISCO DAD3350 dicing saw, the diced chips were glued to ceramic packages (CQFJ-68) and gold wire-bonded with an F\&S BONDTEC Series 53. 
Atomic force microscopy (AFM) and Kelvin probe force microscopy (KPFM) studies were performed on both IGZO-NA and IGZO-A samples. 
These results, presented in supporting information (see Fig.\,S4), confirm the presence of donor-type oxygen-deficiency defects that introduce electronic states near the conduction band in the amorphous IGZO.

The mechanism responsible for the optically-sensitive resistive change in the IGZO devices is the de-ionisation of oxygen-vacancies induced by visible light illumination. 
The device is based on an oxygen-deficient IGZO (OD-IGZO)/oxygen-rich IGZO (OR-IGZO) homojunction, which forms a potential barrier-and-well at the material interface (due to electron diffusion).
The interfacial barrier width (which depends on the ion carrier density) determines tunnelling current.
The optical/electronic band gap changes slightly depending on the oxygen content because oxygen vacancies strongly affect the electronic structure.
OD-IGZO has a higher intrinsic conductivity than OR-IGZO, forming an ohmic contact with the electrode.
When illuminated during the optical WRITE/SET process, the blue light ionises the neutral oxygen vacancies on the oxygen-deficient side, resulting in an increase of electron and oxygen-vacancy ion ($V^{2+}_{O}$) densities. 
The un-trapped electrons move to the conduction band towards the contact (away from the interfacial barrier), resulting in a higher electron density on the oxygen-rich side of the barrier. 
This decreases the barrier width, resulting in an increased tunnelling current. 
During the excitation-free recovery, electrons overcome the barrier (from OR-IGZO to OD-IGZO) due to thermal effects, and are then trapped by the oxygen vacancy ions in the oxygen-deficient side. 
This neutralisation reduces the electron and ion ($V^{2+}_{O}$) density, increasing the barrier width and resulting in a decreased tunnelling current. 
This action of neutralising the ions ($V^{2+}_{O}$) can be induced on a faster time-scale in the RESET process by applying a negative pulse, which provides the electrons with a voltage to overcome the barrier and become trapped \cite{Hu2021,Pereira2024}.

\subsection{Experimental Setup}\label{sec:Setup}

To characterise the electrical and optical response of the ORRAM devices, the optical probe station shown in Fig.\,\ref{fig:Fig1}(a) was assembled. 
Here, 450\,nm blue-light from a $\mu$LED array was collimated (L1) and focused onto the memristive devices via a Nikon N20x-PF microscope objective. 
A 70:30 (Reflection:Transmission) beam-splitter was used in combination with a focussing lens and a IDS U3-3682XLE Rev.1.2 camera, to visualise the chip surface for injection alignment and device selection. 
The initial transmission path of the BS provided an output port for recording reference powers via an optical power meter, as well as the monitoring of the real-time optical input via an avalanche photo-detector and oscilloscope.

The bespoke $\mu$LED chip consisted of a $40 \times 10$ array of $80 \times 80\,\mu$m pixels with a 100\,$\mu$m pitch \cite{Herrnsdorf2015}. 
The $\mu$LEDs were fabricated for 450\,nm blue emission, each producing a measured maximum reference power of 109\,$\mu$W, corresponding to 222\,$\mu$W at the output of MO. 
The $\mu$LED array was flip-chip-bonded to a custom CMOS driver chip \cite{Herrnsdorf2015}, and illumination patterns were programmed into the driver chip through an Opal Kelly XEM3010 field-programmable gate array. 
The CMOS driver had digital input channels for the binary modulation of the entire illumination pattern, or the binary modulation of up to ten independent columns of the array. 
The modulation signal was provided by an AimTTi TGF4242 arbitrary waveform generator (AWG). 

Two wire-bonded chips containing 32 stand-alone IGZO ORRAM devices were used in this work. 
The packaged chips were mounted to and controlled by an ArC ONE board, where the characteristics of the ORRAM devices could be read, and electrically probed. 
Electrical characterisation was conducted using the ArC ONE instrumentation platform alone, whereas optical characterisation was performed using both $\mu$LED optical inputs and electrical ArC ONE reads. 

\section{Conclusion}
In this article, we report the fabrication of optoelectronic IGZO RRAM devices, their electrical and optical characteristics, and the development of an optical injection scheme utilising a bespoke $\mu$LED array for the parallel programming of memristive devices with light. 
We fabricated two-terminal ORRAM devices with a dual-stack active layer of IGZO\textsubscript{Rich} and IGZO, in a 32 stand-alone device configuration. 
We characterise the ORRAM devices both electrically and optically, showing the programming of resistive states without the requirement of electroforming. 
Electrically, the IGZO devices demonstrated an increase in resistance with the application of negative polarity pulses. 
However, under 450\,nm optical stimulation, the devices demonstrated a decrease in their resistive state. 
The optical programming achieved a wide range of resistive states (beyond that of electrical programming) and demonstrated an additional frequency dependence on the incoming optical stimuli. 
Additionally, the optical and electrical characterisation revealed that optically programmed states recovered on a timescale ($>$\,250\,s) much slower than electrically programmed states ($\sim$\,0.1\,s), granting IGZO ORRAM access to a short-term optically-induced fading memory. 
Further, we demonstrated simultaneous photonic and electronic device programming, exploiting counteracting inputs, to achieve an optical SET and electrical RESET behaviour before revealing its potential application in temporal bit pattern encoding. 
In this demonstration, we exploited the fading memory of the ORRAM devices to encode 4-bit patterns with distinct resistive state readouts. 
Finally, by coupling four $\mu$LEDs and four IGZO devices, we demonstrated the simultaneous, parallel and optically-enabled potentiation of multiple ORRAM devices on-chip with a single optical array. 
The optical scheme and the $\mu$LED array demonstrated it was possible to illuminate the selected IGZO devices with 16 different spatial patterns without optical crosstalk or the need for bespoke optical focusing lenses. 
Overall, this work demonstrates a direct pathway to combining array-based optical sources with array-based ORRAM chips for the effective scaling of neuromorphic systems via spatial-multiplexing enabled by photonic technologies. 
This work reveals the possibility of merging electronics and optics for improving the bandwidth and performance of future systems, and highlights the potential to bring neuromorphic memristive technology into applications that directly sense and process signals from the optical domain.          

\medskip
\textbf{Supporting Information} \par 
Supporting information is provided in supplement to this article. The supporting information contains the additional characterisation of an annealed IGZO sample (I-V, R-V, electrical, and optical measurements) and the material surface analysis of both IGZO samples via atomic force microscopy and kelvin probe force microscopy. 

\medskip
\textbf{Author Contributions} \par 
A.A and J.R equal contribution. Idea and Concept: J.H, M.D.D, T.P, and A.H. Device Fabrication: A.T and S.S. Device Test: A.A, J.R, S.S and M.A. System Design: A.A, J.R, and J.H. System Measurement: A.A and J.R. Investigation and Analysis: A.A and J.R. Writing: A.A, J.R, A.T, M.A and S.S. Review and editing: All authors. 

\medskip
\textbf{Funding Statement} \par 
The authors acknowledge support from the EPSRC Project ‘ProSensing’ (reference EP/Y030176/1), EU Pathfinder Open project ‘SpikePro’ (Grant ID 101129904), UK Multidisciplinary Centre for Neuromorphic Computing (reference UKRI982), UKRI Innovation Knowledge Centre (IKC) in Neuromorphic Hardware (Neuroware), and the Royal Academy of Engineering (RAEng) Chair in Emerging Technologies under Grant CiET1819/2/93. 

\medskip
\textbf{Data Availability} \par 
All data underpinning this publication are openly available from the University of Strathclyde KnowledgeBase at https://doi.org/10.15129/6e778ca4-f780-4681-a877-97c1a8793c05.

\medskip
\textbf{Conflicts of Interest Disclosure} \par 
The authors declare no conflicts of interest. 

\medskip

\bibliographystyle{MSP}
\bibliography{references}

\end{document}


\pagestyle{fancy}
\renewcommand{\thefigure}{S\arabic{figure}} 

\rhead{\includegraphics[width=2.5cm]{vch-logo.png}}

\title{Supporting Information - Parallel Spatial Photonic Programming of Optoelectronic IGZO with a compact $\mu$LED Array}

\maketitle

\author{Andrew Adair}
\author{Joshua Robertson*}
\author{Andreas Tsiamis}
\author{Mohamed Awadein}
\author{Spyros Stathopoulos}
\author{Johannes Herrnsdorf}
\author{Martin D. Dawson}
\author{Themis Prodromakis}
\author{Antonio Hurtado}

\dedication{}

\begin{affiliations}
A. Adair, Dr. J. Robertson, Dr. J. Herrnsdorf, Prof. M. D. Dawson and Prof. A. Hurtado\\
Institute of Photonics, SUPA Dept. of Physics, University of Strathclyde, Glasgow, 99 George Street, G1 1RD, UK\\
Email Address: andrew.adair@strath.ac.uk, joshua.robertson@strath.ac.uk, johannes.herrnsdorf@strath.ac.uk, m.dawson@strath.ac.uk, antonio.hurtado@strath.ac.uk

Dr. M Awadein, Dr. A Tsiamis, Dr. S. Stathopoulos, and Prof. T Prodromakis\\
Institute for Integrated Micro and Nano Systems, University of Edinburgh, Edinburgh, Alexander Crum Brown Road, EH9 3FF, UK\\
Email Address: awadein@ed.ac.uk, s.stathopoulos@ed.ac.uk, a.tsiamis@ed.ac.uk, t.prodromakis@ed.ac.uk
\end{affiliations}

\section{Contents}\label{SI-Contents}

In addition to the main manuscript titled `Parallel Spatial Photonic Programming of Optoelectronic IGZO with a compact $\mu$LED Array', this Supporting Information provides details on the electrical and optical characterisation of annealed IGZO (IGZO-A) samples. This includes I-V, R-V, electrical potentiation and depression, and optical potential and depression measurements. Additionally, a detailed material surface analysis, via atomic force microscopy (AFM) and Kelvin probe force microscopy (KPFM), of the fabricated IGZO samples is provided. Each result is discussed in the following sections.

\section{I-V/R-V Characterisation of Annealed IGZO (IGZO-A)}\label{SI-IVIR}

\begin{figure}[h!]
\centering
\includegraphics[width=0.6\linewidth]{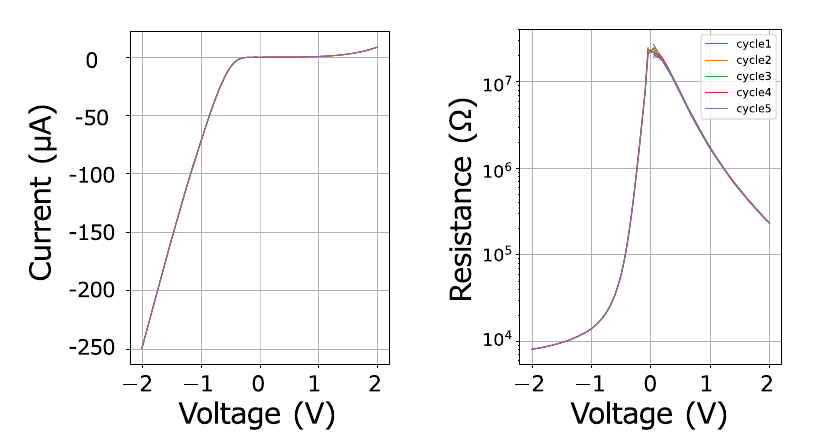}
\caption{\label{fig:FigS1}
    (a) Characteristic I-V and (b) R-V curves (Current and Log Resistance) for the annealed IGZO sample (IGZO-a). Curves cycled 5 times between -2\,V and +2\,V.}
\end{figure}

The motivation behind the characterisation of annealed samples is shown in the R-V curve of the main manuscript (Fig.\,1(b)). When performing the I-V, R-V sweeps of non-annealed samples, the ammeter resolution would cause noisy measurements in the region of 0 to +2\,V. To mitigate this setup limitation, one chip of IGZO devices were annealed at 350\,$^{\circ}$C for 60 minutes. Figure\,\ref{fig:FigS1} shows the I-V and R-V curves of an annealed IGZO device (IGZO-A). Following annealing, the sample demonstrates a significant reduction in intrinsic resistance. The peak resistance is reduced from $10^{10}$ to $10^{7}\,\ohm$, successfully allowing for a non-resolution limited measure of current, resistance and voltage. Unlike IGZO-NA, that required a sweep of $\pm$2.5\,V to read within the resolution of the ammeter, IGZO was successfully measured in a range of $\pm$2\,V. Similar to the IGZO-NA I-V, R-V curves presented in the main manuscript, the annealed devices do not show regions of bistable operation or hysteresis, and in their form-free state have an ohmic linear relationship for negative polarity voltages, and a Schottky contact-induced region of reduced conduction for positive voltage.

\section{Electrical Characterisation of Annealed IGZO (IGZO-A)}\label{Intro}

\begin{figure}[h!]
\centering
\includegraphics[width=1\linewidth]{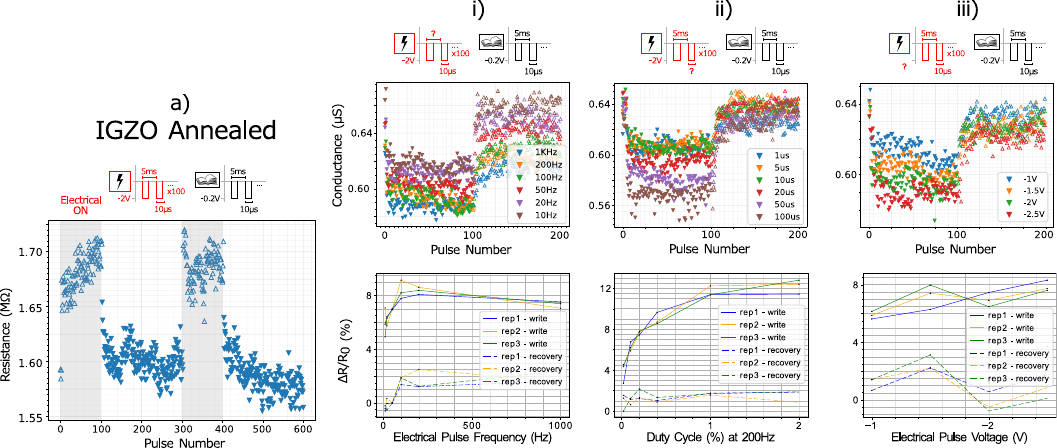}
\caption{\label{fig:FigS2}
    Electrical characterisation of the potentiation-depression property of IGZO-NA with: 
    i) varying pulse frequency, 
    ii) varying pulse width, 
    iii) varying pulse amplitude.
    (Top) conductance ($\mu S$) and (Bottom) relative resistance change (\%) of the devices. 
    The constant parameters are read voltage = -0.2\,V, read interval = 50\,ms, write voltage = -2\,V, pulse width = 10\,us, pulse interval = 5\,ms, pulse count = 100.}
\end{figure}

The electrical characterisation of an IGZO-A device was performed in the same manner as the IGZO-NA characterisation presented in the main manuscript (see Fig.\,2(a)). 
The electrical characterisation measured the potentiation and depression of IGZO-A with 3 varying parameters: pulse frequency, pulse width and pulse amplitude. 
Similar to the main manuscript, 100 excitation pulses were applied, followed by 200 read pulses to observe the recovery of the device (without external stimulation).
Negative polarity pulses were selected for excitation.
For IGZO-A (shown in Fig.\,\ref{fig:FigS2}), the resistance was read using 10\,$\mu$s long pulses of -0.2\,V at an interval of 5\,ms.
A consistent set of starting excitation parameters (-2\,V, 10\,$\mu$s width, 5\,ms interval/200\,Hz frequency) were used across the input pulse studies.
In total three consecutive cycles of potentiation and depression were measured and analysed.
The electrical characterisation results shown in Figs.\,\ref{fig:FigS2}(a)(i-iii) reveal that the influence of electrical WRITE parameters remain consistent across both ORRAM samples.
In agreement with the IGZO-NA samples, the electrical WRITE pulses increase the resistive state of the device and the analysis shows that WRITE frequency has little effect on the final resistive state. 
Overall, the inherent resistance of the IGZO-A sample is reduced, however, both samples produce similar levels of relative resistance change with electrical stimulation.
Seen clearly in the conductance measurements of Figs.\,\ref{fig:FigS2}(a)(i-iii), the noise in the resistance readout of the IGZO-A sample is significantly higher than IGZO-NA.
This creates more overlap in resistive states, reducing the programming resolution for the various WRITE parameters.
Again, this can be seen in the reduced consistency between consecutive cycles in the bottom plots of Figs.\,\ref{fig:FigS2}(a)(i-iii).
These results show that the additional annealing process reduces the overall resistance range of the ORRAM device, but does not equally scale readout noise, producing an electrical WRITE operation with less consistency, control and resolution.
The annealing process was therefore successful in lowering the resistance of the devices, but the electrical programming performance of IGZO-A becomes less desirable than the non-annealed IGZO-NA devices.
The different properties of the annealed sample can be explained through the altered distribution and charge states of oxygen-related defects, as discussed in Section\,\ref{Sect:Mat_analys}.

\section{Optical Characterisation of Annealed IGZO (IGZO-A)}\label{Intro}

\begin{figure}[h!]
\centering
\includegraphics[width=1\linewidth]{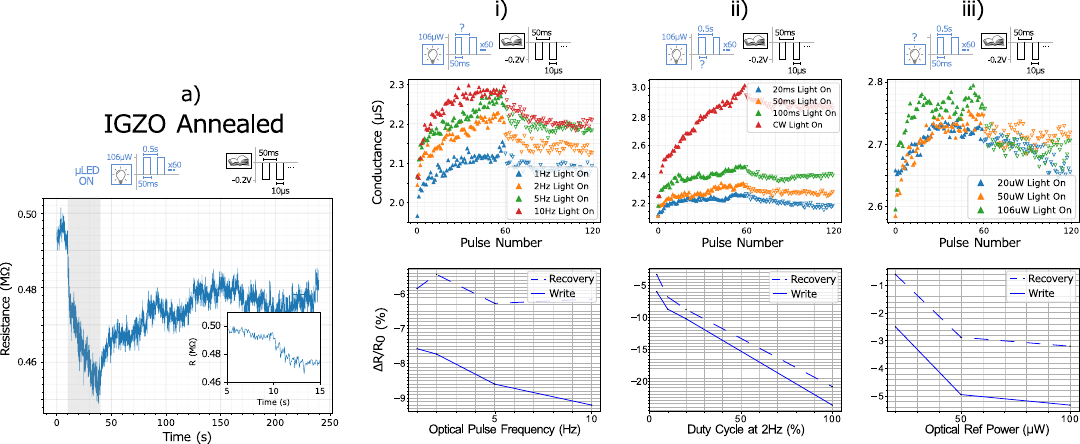}
\caption{\label{fig:FigS3}
    Optical characterisation of the potentiation-depression property of IGZO-NA with: 
    i) varying pulse frequency, 
    ii) varying pulse width, 
    iii) varying optical power.
    (Top) conductance ($\mu S$) and (Bottom) relative resistance change (\%) of the devices. 
    The constant parameters are read voltage = -0.2\,V, read interval = 50\,ms, optical power = 106\,$\mu W$, pulse width = 50\,ms, pulse interval = 500\,ms, pulse count = 60.}
\end{figure}

The optical characterisation of an IGZO-A sample was performed in the same manner as the IGZO-NA characterisation presented in the main manuscript (see Fig.\,3(a)). 
For this characterisation, 60 optical excitation pulses were applied, followed by 400 read pulses to observe the recovery of the device. 
Similar to IGZO-NA measurements in the main manuscript, the optical excitation parameters were varied.
Optical pulse frequency was varied from 1 to 10\,Hz, optical pulse width was varied from 20\,ms to continuous wave (CW), and optical reference power was varied from 4.75 up to 106\,$\mu W$. 
Electrical read pulses of 10\,$\mu$s width and -0.2\,V produced measurements at an interval of 50\,ms. 
Again, a consistent set of starting excitation parameters were used (2\,Hz optical pulse frequency, 50\,ms pulse width, and an optical reference power of 106\,$\mu W$).
Figure\,\ref{fig:FigS3}\,(a) shows that as before, the IGZO-A device has significantly higher noise than IGZO-NA due to the additional annealing process. 
This increased readout noise again influences the consistency, control and resolution of the optically-programmed resistive states, with less definition on the optical potentiation, and a less defined recovery curve.
The parameter analysis in Figs.\,\ref{fig:FigS3}\,(a)(i-iii) also show that overall the optical WRITE operation in the IGZO-A sample produces less resistance change than IGZO-NA.
The programming resolution of IGZO-A is therefore also reduced, resulting in resistive states that overlap within the readout noise (seen clearly in the conductance plots of Figs.\,\ref{fig:FigS3}\,(a)(i-iii)). 
Using a moving average to extract the WRITE and recovery states reveals that despite the overall reduced resistance change, the influence of all 3 optical pulse parameters is similar to IGZO-NA.
Increasing optical WRITE pulse frequency, pulse width and optical power decreases the resistance of the final state, meaning optical programming should consider all 3 parameters.
The annealing process was therefore successful in lowering the resistance of the devices, but the optical programming performance of IGZO-A is significantly reduced, opposing the motivation behind ORRAM and this research investigation. 

\section{Material Surface analysis}\label{Sect:Mat_analys}

\begin{figure}[h!]
\centering
\includegraphics[width=1\linewidth]{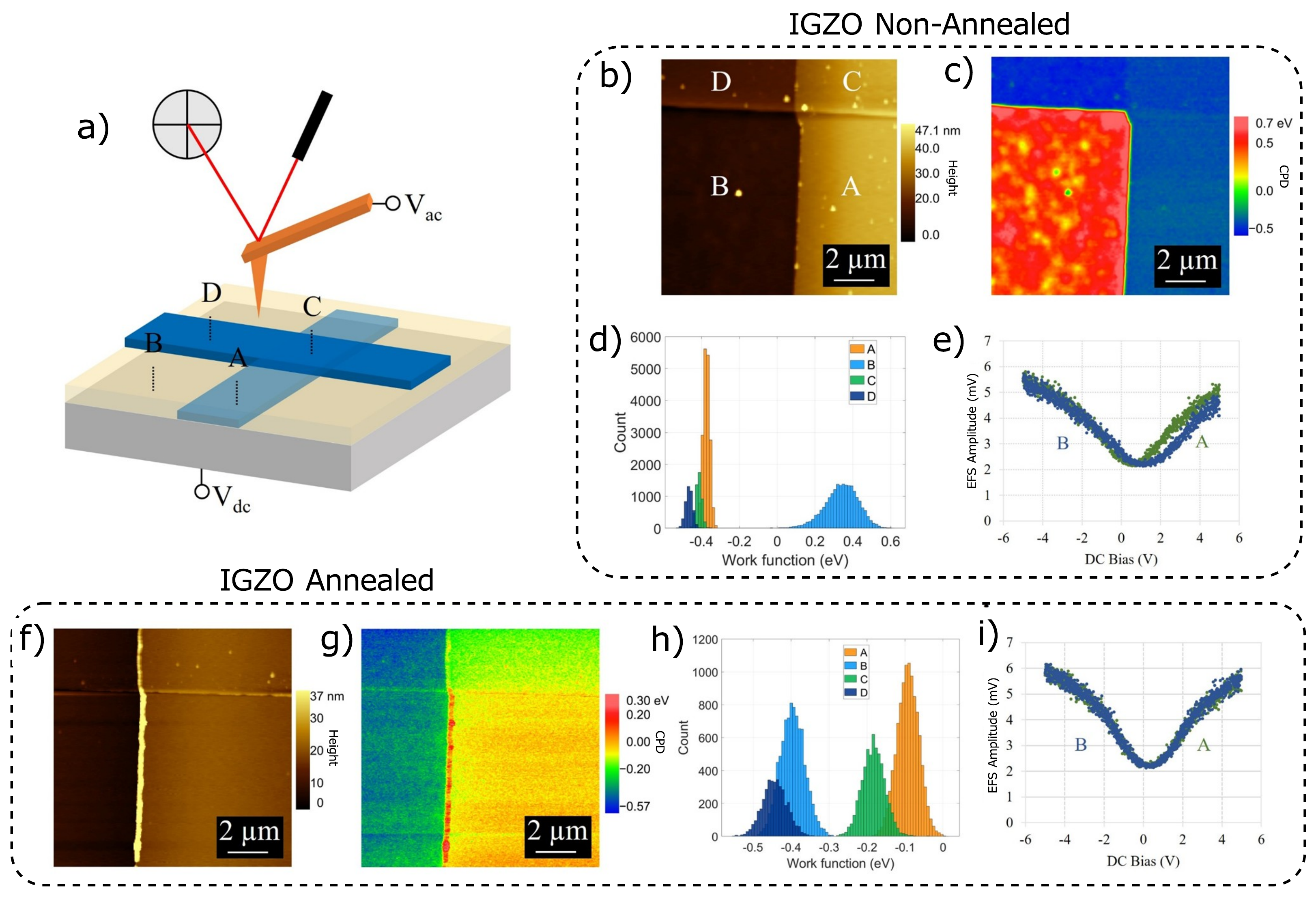}
\caption{\label{fig:Fig-KPFM}
    Surface Measurements of optoelectronic ORRAM Chips. (a) AFM/KPFM measurement configuration and device layout. Regions A–D correspond to distinct stack geometries: A – IGZO layer supported by a bottom ITO electrode; B – IGZO layer without underlying metal; C – ITO/IGZO/ITO stack; D – exposed ITO top electrode, with underlying IGZO layer. (b)-(e) IGZO non-annealed measurements. (b) AFM topography map showing negligible morphological variation between regions A–D. (c) KPFM contact potential difference (CPD) map displaying pronounced contrast between pure IGZO (region B) and electrode-supported regions A, C, D. (d) Corresponding histograms and e) EFS potential-sweep parabolas acquired at fixed positions in regions A \& B, demonstrating a substantial CPD difference between unscreened IGZO (region B) and IGZO in contact with ITO (region A). (f)–(i) Post-annealing IGZO measurements. (f) Topography map confirming preserved surface morphology. (g) KPFM map revealing significantly reduced CPD contrast between regions. (h) Histograms indicating narrowing of CPD distributions after annealing. (i) EFS potential-sweep parabolas at regions A \& B showing reduced CPD separation, consistent with suppression of defect-induced surface band bending and enhanced electrostatic screening in IGZO.}
\end{figure}

The surface characteristics of IGZO-NA and IGZO-A chips were investigated using atomic force microscopy (AFM) and Kelvin probe force microscopy (KPFM) (see Fig.\,\ref{fig:Fig-KPFM}\,(a)). AFM was used to measure the surface topology (Fig.\,\ref{fig:Fig-KPFM}\,(b) \& (f)) and KPFM was used to measure the surface electrostatics (i.e. the contact potential difference (CPD) corresponding to the work-function difference between the probe and the sample surface, Fig.\,\ref{fig:Fig-KPFM}\,(c) \& (g)) \cite{Nonnenmacher1991}. Variations in the measured potential across the device therefore directly reflect differences in surface band bending in the semiconductor \cite{Rosenwaks2004}. For these room-temperature measurements an AC bias V\textsubscript{ac} was applied to the AFM tip, and during KPFM, a DC bias V\textsubscript{dc} was applied to the sample. At room temperature, unsupported IGZO exhibits a pronounced potential shift relative to regions contacted by ITO, indicating stronger surface band bending. Such behaviour is consistent with the presence of donor-type oxygen-deficiency defects that introduce electronic states near the conduction band and modify the Fermi-level position in amorphous IGZO \cite{Noh2011}. 

Oxygen-related defects are widely recognized as the dominant electronic donors that govern carrier concentration and electrostatic potential in this material system \cite{Kamiya2010}. When the oxide is contacted by the conductive ITO electrode, the metallic boundary condition constrains the electrostatic potential through Fermi-level alignment and charge screening at the interface, reducing internal electric fields and suppressing surface band bending \cite{Tung2000}. Independent electrostatic force spectroscopy (EFS) further confirms this interpretation, with a quadratic dependence of force on applied bias yielding parabolic curves whose minima correspond to the local CPD, providing a direct experimental measure of the surface electrostatic potential \cite{Nonnenmacher1991} (see Fig.\,\ref{fig:Fig-KPFM}\,(b)-(e)). Following thermal annealing, the potential contrast between regions decreases significantly (Fig.\,\ref{fig:Fig-KPFM}\,(g)-(i)) while the morphology (Fig.\,\ref{fig:Fig-KPFM}\,(f)) remains unchanged, indicating that the observed electrostatic variations originate from defect-mediated band bending rather than structural differences. Thermal treatment of IGZO is known to modify the distribution and charge state of oxygen-related defects, thereby altering the electronic structure and electrostatic potential landscape of the material \cite{Ryu2010}. This is likely the cause behind the reduced electrical and optical performance of IGZO-A.

\bibliographystyle{MSP}
\bibliography{references}